
\input harvmac

\lref\Affrev{For a review, see I.\ Affleck, ``Conformal Field Theory
Approach to Quantum Impurity Problems'', UBCTP-93-25,
cond-mat/9311054.}
\lref\AFL{N.\ Andrei, K.\ Furuya, and J.\ Lowenstein, Rev. Mod. Phys.
55
(1983) 331; \hfill\break
A.M.\ Tsvelick and P.B.\ Wiegmann, Adv. Phys. 32 (1983) 453.}
\lref\RS{N.\ Reshetikhin and H.\ Saleur, ``Lattice Regularization of
Massive and
Massless Field Theories'', USC-93-020, hep-th/9309135.}
\lref\KR{A.N.\ Kirillov and N.\ Reshetikhin, J. Phys. A20 (1987)
1565,
1587.}
\lref\TS{M.\ Takahashi and M.\ Suzuki, Prog. Th. Phys. 48 (1972)
2187.}
\lref\chered{I.\ Cherednik, Theor. Math. Phys. 61 (1984) 977.}
\lref\GZ{S.\ Ghoshal and A.B.\ Zamolodchikov, ``Boundary State and
Boundary $S$
Matrix in Two-Dimensional Integrable Field Theory'', RU-93-20,
hep-th/9306002.}
\lref\fki{A.\ Fring and R.\ K\"oberle, ``Factorized Scattering in the
Presence of Reflecting Boundaries'', USP-IFQSC/TH/93-06,
hep-th/9304141.}
\lref\fkii{A.\ Fring and R.\ K\"oberle, ``Affine Toda Field Theory in
the
Presence of  Reflecting Boundaries'', USP-IFQSC/TH/93-12,
hep-th/9309142.}
\lref\ghosi{S.\ Ghoshal, ``Bound State Boundary $S$ Matrix of the
Sine-Gordon
Model'',  RU-93-51, hep-th/9310188.}
\lref\ghosii{S.\ Ghoshal, ``Boundary $S$ Matrix of the $O(n)$
Symmetric
Nonlinear Sigma Model'' RU-94-02, hep-th/9401008.}
\lref\sasaki{R.\ Sasaki, ``Reflection Bootstrap Equations for Toda
Field
Theory'', 
 hep-th/9311027.}
\lref\ABBBQ{F.\ Alcaraz, M.\ Barber, M.\ Batchelor, R.\ Baxter
and G. \  Quispel, J. Phys. A20 (1987) 6397.}
\lref\skly{E.K.\ Sklyanin, J. Phys. A21 (1988) 2375.}
\lref\ZandZ{A.B.\ Zamolodchikov and Al.B.\ Zamolodchikov, Ann.  Phys.
120 (1980) 253.}
\lref\AL{I.\ Affleck and A.\ Ludwig, Phys. Rev. Lett. 67 (1991) 161.}
\lref\CarVer{J.\ Cardy, Nucl. Phys. B324 (1989) 581.}
\lref\pkondo{P.\ Fendley, Phys. Rev. Lett. 71 (1993) 2485,
cond-mat/9304031.}
\lref\FS{P.\ Fendley and H.\ Saleur, Nucl. Phys. B388 (1992) 609,
hep-th/9204094.}
\lref\DL{C.\ Destri and J.H.\ Lowenstein, Nucl. Phys. B205 (1982)
369.}
\lref\DNW{G.I.\ Dzapardize,  A.A.\ Nersesyan and P.B.\ Wiegmann,
Phys.
Scr. 27 (1983) 5.}
\lref\DV{C.\ Destri and H.\ de Vega, J. Phys. A22 (1989) 1329.}
\lref\boundrefs{L.\ Mezincescu and R.I.\ Nepomechie, Int. J. Mod.
Phys.
A6 (1991) 5231; Int. J. Mod. Phys. A7 (1992) 565; H.J.\ de Vega
and A.\ Gonzalez-Ruiz preprint LPTHE-93/38.}
\lref\nonstrings{F.\ Woynarovich, J. Phys. A15 (1982)
2985;\hfill\break
O.\ Babelon, H.J.\ de Vega and C.M.\ Viallet, Nucl. Phys. B220
 (1983) 13.}
\lref\EKS{F.\ Essler, V.\ Korepin and K.\ Schoutens, J. Phys A25
(1992)
4115.}
\lref\JW{R.\ Jackiw, G. \ Woo, Phys. Rev. D12 (1975) 1643.}
\lref\KF{V.E.\ Korepin, L.D.\ Faddeev, Th. and Math. Phys. 25 (1965)
147.}
\lref\KKK{V.\ Korepin, Th. and Math. Phys. 34 (1978) 1.}
\lref\dissip{A.O.\ Caldeira, A.J.\ Leggett, Physica 121A (1983) 587.}
\lref\AL{I.\ Affleck, A.\ Ludwig, Nucl. Phys. B360 (1991) 641.}
\lref\CK{C.\ Callan, I.\ Klebanov,  Phys. Rev. Lett. 72 (1994) 1968.}
\lref\FLS{P.\ Fendley, A.\ Ludwig, H.\ Saleur,
``Exact Hall conductance through
point contacts in the $\nu=1/3$ fractional quantum Hall effect'',
preprint USC-94-12.}
\lref\KF{C.Kane, M.Fisher, Phys. Rev. B46 (1992) 15233.}
\lref\FSW{P.\ Fendley, H.\ Saleur, N.P.\ Warner,
``Exact solution of a massless scalar field with a
relevant boundary interaction'', preprint USC-94-10.}

\def\t{\theta}

\def\s{{\cal S}}


\def\coeff#1#2{\relax{\textstyle {#1 \over #2}}\displaystyle}
\def\frac#1#2{{#1 \over #2}}
\def\smallstuff#1{{\relax{\textstyle {#1 }}\displaystyle}}
\def\verysmallstuff#1{{\relax{\scriptstyle {#1 }}\displaystyle}}
\def\inbar{\vrule height1.5ex width.4pt depth0pt}
\def\IC{\relax\,\hbox{$\inbar\kern-.3em{\rm C}$}}
\def\IR{\relax{\rm I\kern-.18em R}}
\font\sanse=cmss12
\def\ZZ{\relax{\hbox{\sanse Z\kern-.42em Z}}}


\def\shalf{\coeff{1}{2}}

%

%

%


\Title{\vbox{\baselineskip12pt
\hbox{USC-94-013}\hbox{hep-th/9408004}}}
{\vbox{\centerline{
The boundary sine-Gordon theory:}\vskip4pt\centerline{classical and
semi-classical analysis}}}

\centerline{H. Saleur$^*$,  S. \ Skorik \ and \ N.P. Warner}
\bigskip \centerline{Physics Department}
\centerline{University of Southern California}
\centerline{University Park}
\centerline{Los Angeles, CA 90089-0484.}
\vskip 1.0cm
We consider the sine-Gordon model on a half-line, with
an additional potential term of the form $-M\cos{\beta\over
2}(\varphi-\varphi_0)$ at the boundary. We compute the classical time
delay for general values of $M$, $\beta$ and $\varphi_0$  using
$\tau$-function methods and show that in the classical limit, the
method of images still works, despite the non-linearity of the
problem. We also perform a semi-classical analysis, and find
agreement with the exact quantum S-matrix conjectured by Ghoshal and
Zamolodchikov.

\vfill
\noindent $^*$ Packard Fellow

\Date{July, 1994}


\newsec{Introduction}

There are a wealth of applications of $1+1$-dimensional quantum field
theories defined upon the half-line.  Such theories generically have
some potential term at the boundary, and are thus often referred to
as ``boundary field theories.''  Amongst other things, they have been
used to describe  dissipative quantum
mechanics \dissip, Kondo effects \AL, and quantum wires
\refs{\KF,\FLS}.
The exemplar of such theories is the sine-Gordon model, whose
Lagrangian is given by:
\eqn\lagr{L={1\over 2}\int_0^\infty
\left[\left(\partial_t\varphi\right)^2 -
\left(\partial_x\varphi\right)^2 +
g\cos\beta\varphi\right]dx ~+~ M\cos
{\beta\over 2}(\varphi(x=0)-\varphi_0) \ .}
The massless limit for $\beta^2=8\pi$  was considered in \CK,
and for generic values of $\beta^2$ in \FSW.

The quantum version of the sine-Gordon model was considered by
Ghoshal
and Zamolodchikov \GZ, who  conjectured its integrability and an
exact S-matrix.
Our purpose here is to discuss these two aspects in the classical and
semi-classical limits, generalizing the well
known bulk analysis  \refs{\JW, \KF, \KKK} to the boundary problem
\lagr.

We establish the classical integrability in the next section by using
the
bulk sine-Gordon theory and the method of images.  Although such a
method might not be expected to work, {\it a priori}, due to the
non-linearity,
it turns out that the boundary condition inherited from \lagr\ can be
realized, for the boundary reflection of a single soliton, by using a
three soliton solution of the bulk problem.  We
compute the classical time delay for the most general values of the
parameters $M$ and $\phi_0$.  The semi-classical analysis is
performed in the third section of this paper.  We  find agreement
with the $\beta\rightarrow 0$ limit of the results in
\GZ.

\newsec{The classical solutions}

For a conformal field theory, or for any linear partial differential
equation, the simplest method of dealing with a boundary is to use
some
version of the method of images \ref\jcardy{J.\ Cardy, Nucl. Phys.
B240 (1984) 514.}.  One of the purposes
of this section is to  describe how this technique can also be
applied to
the classical sine-Gordon equation
on the
semi-infinite interval.
Indeed, we will show that this method
replicates the
soliton scattering from a boundary with the most general boundary
conditions
consistent with integrability \GZ.

\subsec{The $\tau$-functions}

After appropriate rescaling of the field and the coupling constant
we have, for the classical problem
associated with the bulk part of \lagr,  the sine-Gordon equation

\eqn\sgeqn{\phi_{tt} ~-~ \phi_{xx} ~=~ -\sin(\phi) \ .}
where  $\beta\varphi\equiv\phi$. On the interval $[0,\infty)$ it was
argued  in \GZ\ that the most
general boundary conditions consistent with integrability were:
\eqn\intbcs{\del_x \phi |_{x=0} ~=~ M \sin\big( \shalf(\phi - \phi_0)
\big) \big |_{x=0} \ ,}
for some constants $M$ and $\phi_0$, corresponding to the boundary
term in \lagr.

On the infinite interval, $(-\infty,\infty)$,
the classical multi-soliton solution to the sine-Gordon equation
is well known
\ref\ABL{M.\ Ablowitz, H.\ Segur ``Solitons and the inverse
scattering transform'', SIAM Studies in Appl. Math. 1981.}.
  It is usually expressed as:
\eqn\sgsoln{\phi(x,t) ~=~ 4 ~ \hbox{arg} (\tau) ~\equiv~ 4~
\hbox{artan}\bigg(
{{\cal I}m(\tau)  \over {\cal R}e (\tau) }  \bigg) \ ,}
where the $\tau$-function for the $N$-soliton
solution is:
\eqn\taufn{\eqalign{\tau ~=~ \sum_{\mu_j = 0,1} ~  e^{{i \pi \over 2}
( \sum_{j=1}^N \epsilon_j \mu_j)} ~\exp ~\bigg [ - \sum_{j=1}^N
\coeff{1}{2}
\mu_j ~ \Big[  \Big(k_j  + & \frac{1}{k_j} \Big)   x   +
 \Big (k_j - \frac{1}{k_j} \Big) t ~-~  a_j \, \Big ]  \cr ~+~&  2
\sum_{1 \le i<j \le N} \mu_i \mu_j ~ \log\bigg( {{k_i - k_j}
\over {k_i + k_j} } \bigg) \bigg]  \ . }}
The parameters $k_j$, $a_j$ and $\epsilon_j$ have the following
interpretations. The velocity of the $j^{\rm th}$ soliton is given
by:
\eqn\vels{v_j ~=~ \bigg( {{k_j^2 - 1} \over {k_j^2 + 1} } \bigg) \ .}
(Note that $v_j$ is positive for a left-moving soliton.)
The $a_j$ represent the initial positions of each of the solitons,
and
$\epsilon_j = +1$ if the $j^{\rm th}$ soliton is a kink, while
$\epsilon_j = -1$ if it is an anti-kink.  The rapidity, $\theta$,
of the soliton is defined by $k = e^\theta$, and we have normalized
the
soliton masses to unity (in further discussion, the words ``soliton''
 and ``kink'' will be used synonymously).

It is fairly obvious how to get a single soliton solution on
$[0,\infty)$
with either $\phi |_{x=0} = 0$ or $\del_x \phi |_{x=0} = 0$.  One
exploits
the symmetry of \sgeqn\ under $\phi \to -\phi$ and $x \to -x$, and
simply
takes a two soliton solution on  $(-\infty,\infty)$ where one soliton
is a mirror image of the other through $x=0$
\ref\Trull{ R.M.\ Deleonardis, S.E.\ Trullinger and R.F.\ Wallis, J.
Appl. Phys. 51 (1980) 1211-1226;}.
If one does this with a double-kink solution then it satifies the
foregoing Dirichlet condition, while the kink-anti-kink solution
satisfies
the Neumann condition.  It is also not hard to guess how one can go
beyond
this solution:  For $M = \infty$, the boundary condition \intbcs\
reduces
to $\phi |_{x=0} = \phi_0$.  The only way that this can be obtained
from
a multi-soliton solution on $(-\infty,\infty)$ is to put a third,
{\it stationary  } soliton at the origin.

We therefore consider the three soliton solution with $k_1 = k$, $k_2
= 1/k$
and $k_3 = 1$.  That is, we consider:
\eqn\bdrytau{\eqalign{\tau ~=~ 1 ~-~ & \epsilon v^2 e^{ -\coeff{1}{k}
\verysmallstuff{(k^2 +1) x - a}} ~-~ \epsilon_0 \big(\coeff{k-1}{k+1}
\big)^2 e^{- \coeff{1}{2k} \verysmallstuff{ (k +1)^2 x - b}} ~ F(t)
\cr
{}~+~ & i \Big\{ ~ e^{- \coeff{1}{2k} \verysmallstuff{ (k^2 +1) x }}
{}~ F(t)
{}~+~ \epsilon_0 e^\verysmallstuff{- x - b}
{}~-~ \epsilon \epsilon_0 v^2 \big( \coeff{k-1}{k+1} \big)^4
e^{- \coeff{1}{k} \verysmallstuff{(k^2 + k +1) x - a - b}} ~\Big\}
 \ , }}
where we have introduced the shorthand:
\eqn\shorth{\epsilon = \epsilon_1  \epsilon_2 \ , \qquad
\epsilon_0 = \epsilon_3\ , \qquad a = a_1 + a_2\ , \qquad b = a_3 \
.}
The function $F(t)$ is defined by:
\eqn\Fdefn{F(t) ~\equiv~ \epsilon_1 e^{ -\coeff{1}{2k}
\verysmallstuff{(k^2 -1) t - a_1}} ~+~ \epsilon_2 e^{ \coeff{1}{2k}
\verysmallstuff{(k^2 -1) t - a_2}} \ .}
This solution has $\phi = 0$ at $x = \infty$, and for $k>1$ it
describes a
left-moving soliton moving from $x = \infty$ with a right-moving
``image''
starting at $ x = -\infty$.   There is a stationary soliton with
center
located at $x =-b$.  Viewing this as scattering off a boundary at
$x=0$ one can easily see that $a$ is the phase delay of the returned
soliton.
To make this more explicit, observe that $\tau$ has the following
asymptotic behaviour:
\eqn\asymp{\eqalign{\tau(x,t) ~&\sim~ 1 ~+~ i \epsilon ~ e^{-
\coeff{1}{2k}
\verysmallstuff{ [(k^2 +1) x + (k^2 -1)t] - a_1}} \qquad x, -t \to
\infty
\ \ {\rm with} \ \ \smallstuff{{x \over t}} = -
\smallstuff{{k^2 -1 \over k^2 +1}} \ ; \cr
\tau(x,t) ~&\sim~ 1 ~+~ i \epsilon ~ e^{- \coeff{1}{2k}
\verysmallstuff{ [(k^2 +1) x - (k^2 -1)t] - a_2}} \qquad x,t \to
\infty,
\ \ {\rm with} \ \ \smallstuff{{x \over t}} =
+\smallstuff{{k^2 -1 \over k^2 +1}} \ .}}
The problem now is to first show that the $\tau$-function given by
\bdrytau\
provides a solution to the boundary value problem on $[0,\infty)$
defined\
by \sgeqn\ and \intbcs.
Our second purpose is to relate the parameters $a$ and $b$ of
\bdrytau\ to
the parameters $M$ and $\phi_0$ of \intbcs, thereby obtaining the
classical phase delay, $a$, in terms of $M$ and $\phi_0$.

\subsec{The classical phase-delay}

To summarize the computation briefly, one substitutes \bdrytau\ into
\intbcs,
and obtains the constraint:
\eqn\midbc{\eqalign{\big[ {\cal R}e (\tau) \partial_x {\cal I}m
(\tau) ~-&~
{\cal R}e (\tau) \partial_x  {\cal I}m (\tau) \big ]\big|_{x=0} \cr
{}~=&~ M~ \big[ 2  \cos(\shalf \phi_0) ~{\cal R}e (\tau)  {\cal I}m
(\tau) ~-~
  \sin(\shalf \phi_0) ~ \big({\cal R}e (\tau)^2 -  {\cal I}m (\tau)^2
\big)  \big ]\big|_{x=0} \ .}}
One can solve this by brute force substitution for the real and
imaginary
parts of $\tau$, but it is somewhat simpler to find constants
$\alpha$,
$\beta$, $\gamma$ and $\delta$ such that:
\eqn\simpler{\eqalign{\partial_x {\cal R}e (\tau)~ \big |_{x=0} ~&=~
\big[
\alpha~ {\cal R}e (\tau)  ~+~ \beta~{\cal I}m (\tau) \big
]\big|_{x=0}\ , \cr
 \partial_x {\cal I}m (\tau)~ \big |_{x=0} ~&=~ \big[ \gamma~
{\cal R}e (\tau)  ~+~ \delta ~{\cal I}m (\tau) \big ]\big|_{x=0}\ \
.}}
One then finds that \midbc\ can be satisfied if and only if  one has
$\alpha = \delta$, which indeed turns out to be true.
Using this one arrives at:
\eqn\Mphireln{\eqalign{M~ \cos(\shalf \phi_0) ~&=~
- \smallstuff{{ (k^2 +1) \over k}} ~{1 \over \Delta}
\Big\{ \big( 1 ~+~ \epsilon~v^2 e^{-a} \big) ~-~ e^{-2b} \big(
\smallstuff{{k-1 \over k+1}}
\big)^2 \Big[ 1 ~+~ \epsilon~v^2 ~ \big( \smallstuff{{k-1 \over k+1}}
\big)^4 ~e^{-a} \Big] \Big\} \ , \cr
M~ \sin(\shalf \phi_0) ~&=~ - 2~\epsilon_0 ~ \smallstuff{{(k-1)^2
\over k}} ~ e^{-b} ~ {1 \over \Delta} \Big\{  1 ~+~ \epsilon~v^2
{}~ \big(\smallstuff{{k-1 \over k+1}} \big)^2 ~e^{-a} \Big\}
\ \ ,}}
where
\eqn\Deldefn{\Delta ~\equiv~   \big( 1 ~-~ \epsilon~v^2  ~ e^{-a}
\big)
{}~+~ e^{-2b} \big( \smallstuff{{k-1 \over k+1}} \big)^2
\Big[ 1 ~-~ \epsilon~v^2 ~ \big( \smallstuff{{k-1 \over k+1}}
\big)^4 ~e^{-a}  \Big] \ .}

It is algebraically very tedious to invert this relationship.  One
proceeds
by eliminating $e^{-b}$, and then  solving for $a$.  It is very
convenient to
introduce a new parametrization of $M$ and $\phi_0$:
\eqn\newparam{\eqalign{\mu ~\equiv~ M~ \cos(\shalf \phi_0) ~&\equiv~
2~\cosh(\zeta) \cos(\eta) \ ; \cr  \nu ~\equiv~ M~ \sin(\shalf
\phi_0)
{}~&\equiv~ 2~\sinh(\zeta) \sin(\eta) \ ,}}
where $0 \le \zeta < \infty$ and $ -\pi < \eta \le \pi$.
(In the $(\mu,\nu)$-plane the curves of constant $\zeta$ are
ellipses,
while the curves of constant $\eta$ are hyperbolae whose asymptotes
make an angle of $\shalf \phi_0$ with the $\mu$-axis.)
We then find that the phase delay is given by:
\eqn\phasedel{a ~=~ \log\bigg\{ -\epsilon ~\big( \tanh\big(
\coeff{\theta}{2}
\big) \big )^2  \big( \tanh( \theta ) \big )^2  \bigg[
{{\tanh\shalf(\theta +
i \eta) ~ \tanh\shalf(\theta - i \eta) } \over {\tanh\shalf(
\theta+\zeta )
{}~ \tanh\shalf(\theta-\zeta) }} \bigg]^{\pm1} \bigg\} \  .}

There are several things to note about this formula.

\item{(i)} The ambiguity
of the $\pm1$ power comes from solving a quadratic equation, and is
a direct reflection of the fact that \intbcs\ is not invariant under
$\phi \to \phi + 2\pi$ (whereas \sgeqn\ {\it is} invariant under this
shift).
Equivalently, one can flip between the $+$ and $-$ powers by sending
$\phi_0 \to \phi_0 + 2\pi$.

\item{(ii)}  The argument of the log is {\it always} real and
positive.
The discrete parameter, $\epsilon = \epsilon_1 \epsilon_2$, must be
chosen to arrange this.  Hence:
\eqn\epschce{\epsilon = +1 \ \ {\rm for} \ \ - \zeta < \theta < \zeta
\qquad \qquad \epsilon = -1 \ \ {\rm for} \ \ |\theta| > \zeta \ .}
This means that a kink reflects into kink for $- \zeta < \theta <
\zeta $,
and reflects into an anti-kink for $|\theta| > \zeta$.  This is
consistent
with the fact that Dirichlet boundary conditions ($M = \infty$)
cause
a kink to reflect as a kink, whereas Neumann boundary conditions
($M = 0$) cause a kink to reflect as an anti-kink.  Note that
these two domains of parameter space (in which a kink reflects
differently)
are separated from one another by a logarithmic singularity in the
classical phase delay.

\item{(iii)}  The choice of the power $\pm 1$ in \phasedel\
is correlated with the parameter $\eta$ and whether there is a kink,
or anti-kink at the
origin.  Specifically, we have:
\eqn\epszero{\epsilon_0 ~=~ \pm ~sign(\tan(\coeff{\eta}{2})) \ , }
where the $\pm$ is the same as  that of \phasedel.

\item{(iv)}  In the $M \to \infty$, or Dirichlet, limit we see that:
\eqn\Mtoinf{\zeta ~\sim~ \log(M) \ , \qquad \eta ~\sim~ \shalf
\phi_0 \qquad {\rm and} \qquad \epsilon  ~=~ -1 \ ,}
and the phase delay collapses to:
\eqn\simpphasedel{a ~=~ \log\bigg\{\big(
\tanh\big(\coeff{\theta}{2} \big) \big )^2  \big( \tanh( \theta )
\big )^2
  \bigg[ \tanh\shalf(\theta + i \coeff{\phi_0}{2}) ~
\tanh\shalf(\theta -
i \coeff{\phi_0}{2})   \bigg]^{\pm1} \bigg\} \  .}
In this limit \intbcs\ enforces Dirichlet boundary conditions.
It is, however, important to note that
there are {\it two} possible distinct
boundary values: $\phi|_{x=0} = \phi_0$ and $\phi|_{x=0} = \phi_0 +
2\pi$.
 Since the boundary potential is
$-M \cos(\shalf(\phi - \phi_0))$, one sees that $\phi|_{x=0} =
\phi_0$
is stable,  while $\phi|_{x=0} = \phi_0 + 2\pi$ is unstable.

{}From now on we consider only the stable boundary value that
corresponds to the
positive sign in \simpphasedel. Then one has from \epszero
\eqn\epszerosimp{\epsilon_0 ~=~ - ~sign(\tan(\coeff{1}{4}
\phi|_{x=0})) \ . }

It is essential to observe  that we have taken
$\phi_{x=\infty} = 0$,
{\it ab initio}. For different boundary conditions at $x=\infty$,
one should replace $\phi_0$ by $\phi|_{x=0}-\phi|_{x=\infty}$.
This  physical
parameter is defined  mod $4\pi$.  An independent derivation of time
delay for the Dirichlet boundary conditions, along with
further details, are provided in the Appendix.

\subsec{Boundary breather solutions}

To fully understand the semi-classical scattering computation one
also needs another class of classical solutions,
which we call here ``boundary breathers''.   It is well-known that
``breathers''
represent bound states in the soliton-anti-soliton channel in the
bulk
sine-Gordon theory.  In the same way that the classical bulk
breathers  can be obtained from the appropriate solution by analytic
continuation of $\theta$ to
imaginary axis, one might expect that the same procedure would give
{\it
boundary} breathers on the half-line. To see this, we set
$\theta=i\vartheta$
($0 < \vartheta < {\pi\over 2}$) in the formula (A1) (see appendix).
Next, we impose the following conditions for a solution to be
boundary breather:

\item{a)} it should be a real function,

\item{b)} it should have finite energy and

\item{c)} the asymptotic value at $x=+\infty$ must be fixed and equal
to
$2\pi n$
($n$ -- integer number).

The three-soliton  (resp. soliton-anti-soliton-soliton) configuration
satisfies
the first condition provided that $\vartheta<-{\phi_0\over 2}$
(resp. $\vartheta<{\phi_0\over 2}$). However, the other conditions
are satisfied by the soliton-anti-soliton-soliton configuration only,
which
one could have foreseen from the analogy with the bulk theory.
The boundary breather
solution has the form (see figure 2):
\eqn\breath{
\eqalign{ & \phi_b  = 2\pi - \cr & -4\arctan
{2\cot\vartheta\cot{\vartheta\over 2}\sqrt{K}\cot{\phi_0\over 4}
e^{x+x\cos\vartheta}\cos(t\sin\vartheta) + e^{2x\cos\vartheta}K\cot^2
{\vartheta\over 2} + 1 \over 2\cot\vartheta\cot{\vartheta\over
2}\sqrt{K}
e^{x\cos\vartheta}\cos(t\sin\vartheta)+e^x\cot{\phi_0\over 4}(
e^{2x\cos\vartheta}K +  \cot^2{\vartheta\over 2}) },\cr} }
where
$$ K=\cot({\phi_0\over 4} - {\vartheta\over 2})
\cot({\phi_0\over 4} + {\vartheta\over 2}) .$$

So, we have a continuum of classical boundary bound states when
$0<\vartheta<
{\phi_0\over 2}<{\pi\over 2}$ and for other $\phi_0$
according to the
$2\pi$-periodicity. In the quantum theory this continuum shrinks into
a
discrete set of bound states (see next section). Note that in the
limit
$\vartheta\to{\phi_0\over 2}$  the boundary breather \breath\ reduces
to
the ground state, figure 3, and
the phase delay has singularities at $\theta = \pm \coeff{i}{2}
\phi_0$.
An analogous picture of bound states occurs
for the anti-soliton scattering with fixed boundary conditions.

\subsec{General solutions, integrability and B\"acklund
transformations}

\nref\earlywork{E.K. Sklyanin, Funct. Anal. Appl. 21 (1987)
164.}
\nref\earlyworkI{V.O. Tarasov, Inverse Problems 7 (1991) 435.}

Thus far we have only applied the method of images to obtain certain
special classical solutions of the boundary sine-Gordon problem
\lagr.  It is natural to suggest that general solutions of \lagr\ can
be
obtained by similar methods.  This in turn would establish the
classical integrability the boundary problem \lagr.  It is, in fact,
rather straightforward to show that both of these conclusions are
true, at least for the problem \lagr\ with $\phi_0 =0$. The method we
will employ can almost certainly be generalized to problems with
$\phi_0 \ne 0$, and also
has the virtue that it can be used to construct the integrable
boundary potentials for the more general Toda models. A related
approach has
been followed in \refs{\earlywork, \earlyworkI}. The basic
idea is to use the fact that any integrable hierarchy has B\"acklund
transformations: that is, solutions can be mapped into one another by
 non-trivial gauge transformations constructed from the affine Lie
algebra action on the corresponding LAX system.   For the
sine-Gordon equation, the requisite B\"acklund transformation can be
cast in the following form:
\eqn\Backlund{ \partial_u (\phi ~+~ \psi) ~=~ e^\zeta ~
\sin \Big({\phi - \psi \over 2} \Big) \ ;  \qquad
\partial_v (\phi ~-~ \psi) ~=~ e^{-\zeta} ~
\sin \Big({\phi + \psi \over 2} \Big) \ ,}
where $u = x - t$, $v = x + t$, and $\zeta$ is an arbitrary constant
parameter.  The point is that $\phi$ satisfies the sine-Gordon
equation \sgeqn\ if and only if $\psi$ does so as well.  Suppose that
$\psi$ is a solution to sine-Gordon on $[0,\infty)$ satisfying a
Dirichlet boundary condition: $\psi|_{x=0} = 2 \eta$, where $\eta$ is
a
constant. It follows immediately from \Backlund\ that $\phi$ is a
solution to sine-Gordon satisfying \intbcs, where $M$ and $\phi_0$
are given in terms of $\zeta$ and $\eta$ by \newparam.  Thus, if one
can solve the Dirichlet problem, one can solve the more general
problem by a B\"acklund transformation.

Observe that if $\eta =0$, or equivalently $\phi_0 =0$, then the
Dirichlet problem can be solved trivially by method of images: one
gets the solution on the half-line by extending it as an odd
function on the full line.  Thus, the B\"acklund transformation
essentially defines the generalized  method of images.  It is also
by no means an accident that the parameters entering into the
B\"acklund transformation ($\eta$ and $\zeta$) are precisely the
rapidity parameters that turn up in the phase delay \phasedel.   One
should also note that the form of the integrable boundary potential
is given directly by the B\"acklund transformation.  This fact should
easily generalize to Toda systems \ref\durham{E.\ Corrigan, P.E.\
Dorey,
R.H.\ Rietdijk, R.\ Sasaki,
``Affine Toda field theory on a half line'', preprint
1994, hep-th/9404108.}.

B\"acklund transformations, in general, are invertible
transformations on the solution space of an integrable hierarchy.
The simplest forms of them modify the constants of the motion of
solution, and possibily add or subtract a soliton.   One can
certainly find a B\"acklund transformation that will introduce a
stationary soliton into the soliton-soliton solution of sine-Gordon.
As a result, the general three soliton solution employed above can be
obtained from the two soliton solution that is appropriate for the
``trivial'' Dirichlet problem with $\phi_0 = 0$.   We therefore
expect that any solution of the trivial Dirichlet problem can be
mapped onto the generic problem \lagr, thus establishing the
classical integrability.  Here we shall content ourselves with having
explicitly established these results
for $\phi_0 = 0$, and having shown that there is a
direct link between the form of the potential and the structure of
the underlying integrable hierarchy.

\newsec{The semi-classical analysis}

We start the discussion by considering the $\beta^2\rightarrow 0$
limit of the results in \GZ. The (quantum) boundary S matrix
elements are \GZ
\eqn\bsI{\eqalign{S_+^+(\t)\equiv
P_+(\t)&=\cos\left[\xi-(t-1)i\t\right]
R(\t)\cr
S_-^-(\t)\equiv P_-(\t)&=\cos\left[\xi+(t-1)i\t\right]R(\t),\cr}}
and
\eqn\bsII{S_+^-   \equiv Q_+(\t)=S_-^+=Q_-(\t)=
{k\over 2}\sin\left[2(t-1)i\t\right]R(\t) \ ,}
where  $\theta$ is the rapidity of an incoming particle.  The
parameter
t is defined by:
$$
t ~=~ {8\pi\over \beta^2}.
$$
The function $R(\t)$ decomposes as

\eqn\bsR{R(\t)=R_0(\t)R_1(\t),}
where $R_0$ is a normalization factor ensuring unitarity and crossing
symmetry that does not depend on the boundary conditions. The
dependence upon boundary conditions appears in $R_1$, which reads
\eqn\bsRI{R_1(\t)={1\over\cos\xi}\sigma(\eta,\t)\sigma(i\Theta,\t).}
Two of the four parameters  $k,\xi,\eta,\Theta$ are  independent, and
we have
the relations
\eqn\rela{\cos\eta\cosh\Theta={1\over k}\cos\xi,\quad
\cos^2\eta+\cosh^2\Theta=1+{1\over k^2}.}
These parameters are related with $M$ and $\varphi_0$ in an unknown
way. An expression
for the functions $R_0$ and  $\sigma$ involving infinite products of
$\Gamma$-functions is given in \GZ. There are also simple integral
representations:
\eqn\intrep{\sigma(\xi,\t)={\cos\xi\over\cos[\xi-i(t-1)\t]}
\exp\left\{\int_{-\infty}^\infty \quad{dx\over
x}\quad{\sinh(t-1+{2\xi\over\pi})x\over
2\cosh(t-1)x\sinh x}\ e^{i{2\over \pi}(t-1)\t x}\right\},}
 and
\eqn\intrepI{R_0(\t)=\exp\left\{-\int_{-\infty}^\infty\quad{dx\over
x}\quad{\sinh{3\over 2}(t-1)x\sinh({t\over 2}-1)x\over \sinh{x\over
2}\sinh 2(t-1)x}\ e^{i{2\over \pi}(t-1)\t x}\right\}.}
One has  $\sigma(\xi,\t)=\sigma(-\xi,\t)$. The only difference
between the scattering of solitons and anti-solitons arises therefore
from the pre-factor $\cos[\xi\mp i(t-1)\t]$ in \bsI.

For simplicity we consider only the limit in which
$M\rightarrow\infty$. One can identify
the corresponding values $k=0$ and $\Theta=\infty$ easily since,
at these values,
the topological charge $Q={\beta\over 2\pi}\int_0^{\infty}
\partial_x\varphi dx$ is conserved
and therefore the amplitudes $Q_\pm$ must vanish. One has also
$\eta=\xi$ so
\eqn\fix{R_1(\t)={1\over\cos\xi}\sigma(\xi,\t).}

 Consider now the quantum boundary S matrix at the leading order in
${1\over\beta^2}$ as
$\beta\rightarrow 0$. The computation is most easily  done by  using
the integral representation given above, and evaluating the integrals
explicitly by the residue theorem. This provides convergent
expressions where the $\beta\rightarrow 0$ limit can be investigated
term by term. To get non-trivial results one must scale $\xi$ as
$$
{\beta^2\over 8\pi}\xi\rightarrow\hat{\xi}
$$
We  find then

\eqn\semiI{ P_{\pm}(\t)=\exp\left(\pm  {8i\pi
\hat{\xi}\kappa\over\beta^2} + {8i\pi
|\hat{\xi}|\kappa\over\beta^2}\right)
{\s(\t;0)[\s(2\t;0)]^{1/2}\over
[\s(\t;\hat{\xi})\s(\t;-\hat{\xi})]^{1/2}},}
where
\eqn\sdef{\s(\theta; y)=\exp\left({8i\over\beta^2}\int_0^\t
dv\ln\tanh^2{v+iy\over 2}
\right),}
$\s(\t;0)$ being the semi-classical approximation to the bulk soliton-soliton
S-matrix \JW\ (in the following we denote it also as $\s(\t)$),
and $\kappa={\rm sign }\theta$ (in the following  we  assume  that
$\theta>0$).

Before discussing the relation between formula \semiI\ and the
classical
computations of the preceding section, it is useful to comment on the
 bound states of the quantum theory. Poles of the $R_0$
term are
associated with breathers and do not correspond to the boundary
(new)
bound states.  The latter correspond rather to poles  of the
$\sigma(\xi,\t)$ term which are located in the physical strip ${\rm
Im}\t\in [0,\pi/2]$. By inspection of the
 $\Gamma$-product expression  \GZ\
one finds two families  of  poles of $\sigma$:
\eqn\poles{\eqalign{\t_{n,l}^{(1)}&=-i{\pi\over t-1}\left(n+{1\over
2}\right)\pm i{\xi\over t-1}-2il\pi\cr
\t_{n,l}^{(2)}&=i{\pi\over t-1}\left(n+{1\over 2}\right)\pm
i{\xi\over t-1}+(2l+1)i\pi,\cr}}
where $n,l$ are integers. Let us restrict to $\xi>0$. Then the only
physical poles are those that correspond to $+$ sign in $\t^{(1)}$
and $-$ sign in
$\t^{(2)}$. The  first pole that enters the physical strip
(from the bottom) is
$\t^{(1)}_{0,0}$ for $\xi\geq {\pi\over 2}$. The  number of poles of
$\sigma$ in the physical strip increases monotonically with $\xi$ for
$\xi$ small enough, and as $t$ gets large it becomes simply of the
form $\xi/\pi$. These poles are cancelled by the cosine term in $P_-$
and therefore appear only in the $P_+$ amplitude. As $\xi$ reaches
the value $\xi={4\pi^2\over\beta^2}$ the poles densely fill the
interval
$0<{\rm Im}\theta<{\pi\over 2}$ and a pole at $\theta=i\pi/2$
appears corresponding to the emission of a zero momentum soliton.  As
argued in \GZ\ this corresponds to a change in the ground state of
the system. For $0<\varphi_0<{\pi\over \beta}$ the ground state is
$\varphi\rightarrow 0$ at infinity but for
${\pi\over\beta}<\varphi_0<3{\pi\over\beta}$ it is
$\varphi\rightarrow {2\pi\over\beta}$ at infinity. Therefore the
value
$\xi={4\pi^2\over\beta^2}$ corresponds to $\varphi_0={\pi\over\beta}$
and it is the upper  physical value for $\xi$: as
$\varphi_0$ varies arbitrarily, $\xi$ never gets larger than
${4\pi^2\over\beta^2}$ and the set of poles $\t_{n,l}^{(2)}$ never
enters the physical strip.
Observe that there are bound states in the quantum theory when in the
classical
case the kink sitting in the middle and the incoming one are of
opposite topological charges in complete agreement with the
discussion of boundary
breathers in section 2.

To establish the relation between \semiI\ and
\simpphasedel, first recall \JW\ the general relation between the
quasi-classical scattering phase shift, $\delta(\theta)$,
and the classical phase shift, $a(\theta)$~:
\eqn\sergei{\delta(\theta)={\rm constant }+{m\over 2}\int_0^\theta
a(\eta)d\eta,}
Here, $m$ is the  classical mass of the particles involved in the
scattering.
Using the ``semi-classical Levinson
theorem'' to determine the constant in this formula we deduce,
using \simpphasedel, the relation
\eqn\levin{P_+(\t)\equiv e^{2i\delta(\t)},}
where
\eqn\levini{2\delta(\theta)=2n_B\pi+{8\over\beta^2}
\int_0^{\theta}d\eta\log
{\tanh^2\eta \tanh^2{\eta\over 2} \over \tanh{1\over 2}(\eta+i{\beta
\varphi_0\over 2}) \tanh{1\over 2}(\eta-i{\beta\varphi_0\over 2})}, }
and $n_B$ is the number of bound states. Since according to the
preceeding discussion $n_B={\xi\over\pi}={8\hat{\xi}\over\beta^2}$
for
$0<\xi<{4\pi^2\over\beta^2}$ as $\beta\rightarrow 0$ we see that
formula \semiI\ is in complete agreement with the classical phase
shift
\simpphasedel\ and that $\xi$ is proportional to $\phi_0$. In the
region $-{4\pi^2\over\beta^2}<\xi<0$ there are
no physical poles for $P_+$ and $n_B=0$.
 A similar  discussion can be carried out for
$\xi<0$ and $P_-$. As $\beta\rightarrow 0$ the number of physical
poles of $P_-$ varies as
$n_B=-{\xi\over\pi}=-{8\hat{\xi}\over\beta^2}$ and again we have
agreement with the semi-classical Levinson theorem.

 From comparison of \semiI\ and the classical phase shift we see that
$\xi$ and $\varphi_0$ are related linearly as
$\xi={4\pi\over\beta}\varphi_0$. This leads correctly to  the
emission of a zero momentum soliton with the ground state
degeneracy as discussed in \GZ. This linear relation  can hold only
for $|\xi|<{4\pi^2\over\beta^2}$. Beyond that value the ground state
changes. To  compare the quantum result with the classical phase
shift we must then  correlate appropriately  the value of $\phi$ at
infinity in the  latter.  The net effect is to
replace $\phi_0$ by $\phi_0-2\pi$. Eventually, the variation of $\xi$
with
$\varphi_0$ is
therefore
\eqn\variat{\xi={4\pi\over\beta}\left(\varphi_0-{2\pi\over\beta}{\rm
Int }\left[{\beta\varphi_0\over 2\pi}+{1\over 2}\right]\right),}
as illustrated in figure 4. It is very likely that \variat\
is exactly true for finite $\beta$ as well.

We can finally recover \semiI\ without appealing to our knowledge of
quantum boundary bound states. For this one has to evaluate the
action for the classical configuration.
Following the discussion in \JW\ we
have,
\eqn\phase{2\delta(\theta)=C(\varphi_0)+{8\over\beta^2}
\int_0^{\theta}d\eta\log
{\tanh^2\eta \tanh^2{\eta\over 2} \over \tanh{1\over 2}(\eta+i{\beta
\varphi_0\over 2}) \tanh{1\over 2}(\eta-i{\beta\varphi_0\over
2}) },\quad {\rm if} \quad
|\beta\varphi_0|<\pi,}
where we used the soliton mass
$m={8\over\beta^2}$, and $\delta$ satisfies the differential equation
\eqn\alta{\delta(\theta) ~-~ \tanh\theta ~\delta'(\theta)
{}~=~ \int_0^{+\infty}dt
\left( \int_0^{+\infty}dx\dot{\varphi}^2 -
8\sinh\theta\tanh\theta\right).}
We restrict to the situation where we send in a kink, and $\varphi_0$
is
positive and  so there is an anti-kink at the origin (see figure 1).
In this case there are quantum boundary bound states.
It is difficult to
perform the double integral for the three-soliton solution
explicitly because of the cumbersome expression for the integrand (A1).
One might hope that in order to find the $\theta$-independent piece of the
phase shift it is sufficient to evaluate both sides of \alta\ in the
limit as  $\theta\rightarrow +\infty$ or $\theta\to 0$. However both of these
limits do not seem to be helpful, because due to the non-uniform
convergence of the
right hand side  of \alta\ it is not allowed to interchange such a  limit
with the integration. We evaluated the right hand side  of \alta\ (with fixed
$\t$)
numerically for
several different values of $\phi_0$   using  Mathematica.
Combining \phase\ and \alta, we obtain an estimate for $C(\varphi_0)$, in
 agreement with  $C(\varphi_0)={8\pi\over\beta}\varphi_0$
with accuracy $0.1\%$. This result
was checked for different values of $\t$. We therefore obtain agreement
with the semi-classical Levinson theorem.

It is also interesting to investigate the cross-unitarity relation of
\GZ\ in this semi-classical limit. In general we have
\eqn\bdrcrossi{P_+\left({i\pi\over 2}-\t\right)=S_T(2\t)P_-\left(
{i\pi\over 2}+\t\right)+S_R(2\t)P_+\left({i\pi\over 2}+\t\right),}
where $S$ are the bulk S matrix elements. In the semi-classical limit
we are considering, the reflection amplitude  $S_R$ vanishes.
In this same limit we use \semiI\ to find
\eqn\ratio{\eqalign{{P_+\left({i\pi\over 2}-\t\right)\over P_-\left(
{i\pi\over 2}+\t\right)}=&{{\cal S}\left({i\pi\over 2}-\t\right)\over
{\cal S}\left({i\pi\over 2}+\t\right)}\left[{{\cal S}\left({i\pi\over
2}-\t+i\hat{\xi}\right) {\cal S}\left({i\pi\over
2}-\t-i\hat{\xi}\right)\over{\cal S}\left({i\pi\over
2}+\t+i\hat{\xi}\right) {\cal S}\left({i\pi\over
2}+\t-i\hat{\xi}\right)}\right]^{-1/2}\cr
&\left[{{\cal S}(i\pi-2\t)\over {\cal
S}(i\pi+2\t)}\right]^{1/2} \cr}.}
We simplify this expression by  using the following relations deduced
from unitarity and crossing symmetry at this order
\ref\ZZ{A.B.\ Zamolodchikov, Al.B.\ Zamolodchikov, Ann. Phys. 120
(1979), 253.}
\eqn\smatrels{\eqalign{S\left({i\pi\over
2}+\t\right)&=S_T\left({i\pi\over 2}-\t\right)\cr
S(\t)&={1\over S(-\t)}\cr
{\cal S}(\t)&=e^{-i\kappa{8\pi^2\over\beta^2}}{\cal S}_T(\t),}}
to obtain \bdrcrossi\ indeed.

\newsec{Conclusions}

Although the method of images works nicely in the classical theory,
it
does not seem to extend to the quantum case: we have not been able to
recast the boundary S matrix of \GZ\ as a product of bulk S-matrix
elements. A
related phenomenon is the non-trivial structure of the boundary
S-matrix with
one-loop corrections (the semi-classical expressions being the tree
approximation). Recall
that in the bulk one has
\eqn\sboloop{{\cal
S}_1(\theta;y)=\exp\left({8i\over\beta'^2}\int_0^\theta dv\ln\tanh^2
{v+iy\over 2}\right),}
where
\eqn\betaprime{\beta'^2\equiv \beta^2{8\pi\over
8\pi-\beta^2}={8\pi\over t-1}.}
Scaling
$$
{\beta'^2\over 8\pi}\xi\rightarrow\hat{\xi}_1,
$$
one finds the following next-to-leading correction to the boundary
S-matrix in the Dirichlet problem:
\eqn\sbdroloop{P_\pm(\theta)\approx\exp\left(\pm \kappa
{8i\pi\hat{\xi}_1\over
\beta'^2}+\kappa {8i\pi|\hat{\xi}_1|\over
\beta'^2}\right){{\cal S}_1(\theta)[{\cal S}_1(2\theta)]^{1/2}\over
[{\cal S}_1(\theta;-\hat{\xi}_1){\cal
S}_1(\theta;\hat{\xi}_1)]^{1/2}}
\left[\tanh\left({\theta\over 2}-i{\pi\over 4}\right)\right]^{1/2}.}
In addition to the usual replacement of $\beta^2$ by $\beta'^2$ we
see the appearance of an entirely new factor involving the
square-root of a
hyperbolic tangent.

It would be interesting to
investigate further the structure of boundary bound states, most
particularly in the quantum theory. Also, besides the
semi-classical limit, one could also take the non-relativistic limit
of the
S-matrix.  The appropriate candidate for the resulting
non-relativistic
 potential would  be probably
the exactly-solvable P\"oschl-Teller model on half-line
\ref\Teller{G. P\"oschl, E. Teller, Z. Physik 83 (1933) 143.}:
\eqn\PT{ U(x) = U_0 \left( {\mu(\mu-1)\over\sinh^2\alpha x} -
               {\nu(\nu-1)\over\cosh^2\alpha x} \right),}
where $\alpha$, $\mu$, $\nu$ and $U_0$ are some constants,
related
to the parameters $\beta$ and $\varphi_0$ of sine-Gordon model.
Note that \PT\ is simply a superposition of the repulsive
hard-core kink-kink potential (first term) and the kink-antikink
potential
(second term), which are known from the bulk theory \ZZ. This problem will
be studied elsewhere \ref\KS{A.Kapustin, S.Skorik, in preparation.}.

\bigskip

\leftline{\bf Acknowledgements:} We would like to thank P.\ Fendley
 for interesting
discussions and A.Kapustin for the help with numerical
computation and many useful remarks.
  N.W. would like to thank the Theory Division at CERN
for its hospitality while this work was completed.   This work was
supported by the Packard foundation, the National Young Investigator
program (NSF-PHY-9357207) and the DOE under grant number
DE-FG03-84ER40168.

\bigskip

\leftline{\bf Appendix}

We give here the explicit general form of classical solutions and
directly extract
the time delay \simpphasedel\ in the limit $M=\infty$.

In the center of mass reference frame the solution to \sgeqn\ reads:

$$\phi=\mp 4\arctan{2e^{x{\rm ch}\theta - a_1} {\rm ch}
(t{\rm sh}\theta) \pm e^{x - a_3} \mp e^{x(1+2{\rm ch}\theta) - 2a_1
-
a_3 +
2\log(u^2v)}\over  2e^{x(1+{\rm ch}\theta) - a_1 - a_3 +
2\log(u)}{\rm ch}
(t{\rm sh}\theta) \pm  e^{2x{\rm ch}\theta - 2a_1 + 2\log(v)} \mp 1
 },
\eqno(A1)$$
where $u=\tanh({\theta\over 2})$, $v=\tanh(\theta)$, and we
have set
$a_1=a_2$, which means that the solution is invariant under
the
transformation $t\rightarrow-t$.
 The upper (resp. lower) sign corresponds to the situation when
a stationary soliton (resp. anti-soliton) is used to adjust the value
of field at
the boundary. Solution (A1) refers to the case when the incoming
and outgoing particle is the soliton with asymptotic value
 $\phi=2\pi$
at $x=+\infty$.

Let us represent (A1) in the form of rational function of variable
${\rm ch}(t{\rm sh}\theta)$:
$$ \phi(x, t) = 4\arctan {r_2 + s_2{\rm ch}(t{\rm
sh}\theta)\over r_1 + s_1 {\rm ch}(t{\rm sh}\theta) }. $$
The condition $\phi(x=0, t)=\phi_0$ implies that
$$ {r_2\over r_1}={s_2\over s_1}=\tan\left({\phi_0\over
4}\right),\eqno(A2)$$
from which follows immediately
$$ a_3 = \log\left(\mp u^2\tan{\phi_0\over 4}\right).$$
For the argument of logarithm to be positive one should take
$\phi_0>0$
with the stationary anti-soliton and $\phi_0<0$ with the stationary
soliton.
This is illustrated in figure 1. Further, we obtain from (A2)
$$ 2a_1=\log \left[ u^2v^2 { u^2 + \tan^2({\phi_0\over 4}) \over
1 + u^2\tan^2({\phi_0\over 4})}\right], \eqno(A3) $$
which agrees  with \simpphasedel\ where $a=2a_1$. Note that
the time delay, obtained by (A3),
is in fact always  a time {\it advance} in both the attractive and
repulsive
cases. For the same value of $\phi_0$ the
time delay
for the soliton that lives on the left half-line $x<0$ is not the
same as
that of the right half-line soliton
(except for $\phi_0=\pm\pi$). It differs by the sign of
power factor in
the formula \simpphasedel. The position of the ``left'' soliton
is not the exact
 mirror image of the ``right'' soliton for generic $\phi_0$.

\listrefs

\centerline{\bf Figure captions}
\smallskip
\noindent Figure 1: a solution of the classical sine-Gordon equation
with  fixed boundary
conditions. For  $\phi|_{x=0}=\phi_0={3\pi\over 8}$
the solution is
constructed
out of a left-moving soliton, its right-moving image at $x<0$ and
the stationary
anti-soliton in the middle (upper graph).
For $\phi|_{x=0}=\phi_0=-{3\pi\over 8}$
the solution is built out of three solitons (lower graph).

\smallskip

\noindent Figure 2: a boundary bound state (boundary breather) for
the values
 $\theta={i\pi\over 6}$ and $\phi_0={3\pi\over 4}$ at four different instants
of  time.

\smallskip

\noindent Figure 3: two possible ground state configurations for
$\phi_0={3\pi\over 4}$.
The configuration with asymptotic
behaviour $\phi\to 0$ at infininty has lower energy than the other
one.

\smallskip

\noindent Figure 4: variation of $\xi$ with $\phi_0$.

%
%
%
%
%
%
%

\end